\documentclass[showpacs,preprintnumbers,superscriptaddress]{revtex4}
\usepackage{CJK}
\usepackage{amsmath,amssymb,graphicx,bm}
\begin{document}

\title{Quantum hoop conjecture and a natural cutoff for vacuum energy of a scalar field}

\author{Rongjia Yang \footnote{Corresponding author}}
\email{yangrongjia@tsinghua.org.cn}
\affiliation{College of Physical Science and Technology, Hebei University, Baoding 071002, China}
\affiliation{Hebei Key Lab of Optic-Electronic Information and Materials, Hebei University, Baoding 071002, China}
\affiliation{State Key Laboratory of Theoretical Physics, Institute of Theoretical Physics, Chinese Academy of Sciences,
Beijing 100190, China}

\begin{abstract}
We propose here a quantum hoop conjecture which states: the de Broglie wavelength of a quantum system cannot be arbitrarily small, it must be larger than the characterized Schwarzschild radius of the quantum system. Based on this conjecture, we find an upper bound for the wave number (or the momentum) of a particle, which offers a natural cutoff for the vacuum energy of a scalar field.
\end{abstract}


\maketitle

\section{Introduction}
In the past years, a lot of independent cosmological observations, such
as supernova (SN) Ia at high redshift \cite{Riess:1998cb, Perlmutter:1998np}, the cosmic microwave background
(CMB) anisotropy,\cite{Spergel:2003cb, Ade:2013zuv} and large-scale structure \cite{Tegmark:2003ud}, have confirmed that the Universe is undergoing an accelerated
expansion. In the framework of general relativity, an unknown energy component, usually called dark energy, has to be introduced to explain this phenomenon. The simplest
and most theoretically appealing scenario of dark energy is the
vacuum energy which is about $\rho_{\rm ovac}\sim (10^{-3} \rm{eV})^4=10^{-8} \rm{ergs/cm}^3$ matched from observational data. However, this model is confronted with a very difficult problem--cosmological constant problem \cite{Weinberg:1988cp, Carroll:2000fy, Martin:2012bt, Padilla:2015aaa, Padmanabhan:2002ji} (may suffer from age problem as well \cite{Yang:2009ae}). To briefly illustrate this issue, we consider, for example, the vacuum energy density of a scalar field. It is well known that the total vacuum energy density of a scalar field with mass $m$ is quartically divergent in the ultraviolet (UV)
%%%%%%%%%%%%%%%
\begin{eqnarray}
\label{vac}
\rho_{\rm tvac}=\langle0|\hat{\rho}_{\rm tvac}|0\rangle=\int dk \frac{k^2\hbar}{4\pi^2c^2}\sqrt{k^2c^2+m^2c^4/\hbar^2}.
\end{eqnarray}
A usually used regularisation for this divergence is to artificially take a UV cutoff. But if we take different UV cutoffs, such as electroweak scale, grand unification scale, or Planck scale, we can get different values of vacuum energy density. Furthermore the differences between these values are huge, see for example, taking electroweak scale, we get $\rho_{\rm tvac}\sim (10^{11} \rm{eV})^4=10^{48} \rm{ergs/cm}^3$; taking Planck scale, we have $\rho_{\rm tvac}\sim (10^{27} \rm{eV})^4=10^{112} \rm{ergs/cm}^3$. The ratio of theoretical to observational value of the vacuum energy ranges from $10^{56}$ to $10^{120}$. This is the well known cosmological constant problem \cite{Weinberg:1988cp, Carroll:2000fy, Martin:2012bt, Padilla:2015aaa, Padmanabhan:2002ji}. Which scale we should take is still an open problem. Can we find a UV cutoff from fundamental laws of physics?  This is the major issue we will consider in this letter.

Here, combining with quantum and black hole physics, we find an upper bound for the wave number of a quantum particle, which gives a natural cutoff for the vacuum energy of a scalar field.

The rest of the paper is organized as follows. In next section, we will present the upper limit of the wave number from quantum and black hole physics and consider a cutoff of the vacuum energy of scalar field. Finally, we will briefly summarize and discuss our results in section III.

\section{Upper bound for wave number and a natural cutoff for vacuum energy}
For a quantum particle with mass $m$, the de Broglie relation reads $E=\hbar \omega,~~~~\vec{p}=\hbar\vec{k}$. According to the mass-energy relation in special relativity, the total energy of a particle is $E^2=p^2c^2+m^2c^4$. Combining the de Broglie relation and the mass-energy relation, then we have
\begin{eqnarray}
\label{medeb}
E^2=\hbar^2k^2c^2+m^2c^4.
\end{eqnarray}
This equation indicates that $E\longrightarrow \infty$ for $k\longrightarrow \infty$. A natural question rises: is this result reasonable? In other words, because $\omega=\sqrt{k^2c^2+m^2c^4/\hbar^2}$, the question can also be stated as: can a particle oscillate arbitrarily fast (or, can the de Broglie wavelength of a particle be arbitrarily small)? If we take into account the effect of gravitation, the answer may be not.

Think of black hole physics, a system with total energy $E$ has an effective mass $E/c^2$, so it will be characterized with a Schwarzschild radius which is given by
\begin{eqnarray}
\label{sr}
r_{\rm c}=\frac{2G}{c^3}\sqrt{\hbar^2k^2+m^2c^2}.
\end{eqnarray}
The hoop conjecture in black hole physics states: if matter is enclosed in sufficiently small region, then the system should collapse to a black hole \cite{1972Thorne,1991Flanagan}. Similar assumptions were also suggested in \cite{Hong:2004rq,Aste:2004ba,Japaridze:2015tva}: for example, it argued that the energy of a system of size $L$ must have an upper bound not to collapse into a black hole \cite{Hong:2004rq}. Here we generalize the hoop conjecture to the quantum case: the de Broglie wavelength of a quantum system can not be arbitrarily small, it should be larger than the characterized Schwarzschild radius of the quantum system. This can be called quantum hoop conjecture.

This quantum hoop conjecture can get supports from earlier works in literatures. Possible connection between gravitation and the fundamental length was discussed in \cite{Mead:1964zz}. From quantum mechanics and classical general relativity, it was shown in \cite{Calmet:2004mp, Calmet:2005mh} that any primitive probe or target used in an experiment must be larger than the Planck length, which implies a device independent limit on possible position measurements. Researches from string theory, black hole physics, and quantum gravity also predict that there exists a minimum measurable length scale which is approximately equivalent to the Planck length $l_{\rm p}$ \cite{Konishi:1989wk, Maggiore:1993rv, Hossenfelder:2012jw, Garay:1994en, Yang:2009vf}. Based on these researches, we can conclude that the de Broglie wavelength of any quantum system must not be less than the minimum length scale. This conclusion is consistent with the quantum hoop conjecture proposed here: the de Broglie wavelength of a quantum system should be larger than its characterized Schwarzschild radius. In \cite{Casadio:2013uga}, a quantum hoop conjecture was also suggested by constructing the horizon wave-function for quantum mechanical states representing two highly boosted non-interacting particles, which is different from the conjecture we proposed here.

The quantum hoop conjecture suggested here provides: $\lambda>r_{\rm c}$, which gives an upper bound for the wave number
\begin{eqnarray}
\label{wn}
k=\frac{2\pi}{\lambda}<\frac{2\pi}{r_{\rm c}}=\frac{\pi c^3}{G}\left[\hbar^2k^2+m^2c^2\right]^{-\frac{1}{2}}< \frac{\pi c^3}{Gk\hbar}.
\end{eqnarray}
It is easy to get
\begin{eqnarray}
\label{kb}
k<\sqrt{\pi}l^{-1}_{\rm p},
\end{eqnarray}
where $l_{\rm p}=\sqrt{G\hbar/c^3}$ is the Planck length. This bound only holds in the observer's reference frame. Bound (\ref{kb}) also gives an upper limit for the momentum of the particle: $p<\sqrt{\pi}\hbar l^{-1}_{\rm p}$. Obviously, the wave number of a massive particle is less than that of a massless particle.

As an application, we apply the bound for the wave number (\ref{kb}) to the vacuum energy of a scalar field. For a quantum particle of a scalar field, there are three freedoms for oscillation: $k=\sqrt{k^2_{\rm x}+k^2_{\rm y}+k^2_{\rm z}}$. So we have $k<\frac{2\sqrt{3}\pi}{r_{\rm c}}<\sqrt{\sqrt{3}\pi}l^{-1}_{\rm p}$ which offers a natural cutoff for the vacuum energy of a scalar field (\ref{vac})
\begin{eqnarray}
\label{vac1}
\rho_{\rm tvac}=\langle0|\hat{\rho}_{\rm tvac}|0\rangle=\int_0^{k_{\rm max}} dk \frac{k^2\hbar}{4\pi^2c^2}\sqrt{k^2c^2+m^2c^4/\hbar^2}.
\end{eqnarray}
For $k\gg m$, integration (\ref{vac1}) is approximatively equivalent to $3\hbar/(16cl^{4}_{\rm p})$ which closes to the value obtained by taking the Planck scale cutoff. Also based on black hole physics, a cutoff for vacuum energy of a scalar field was found in \cite{Culetu:2004ta}.

\section{Conclusions and discussions}
In this letter, we suggested a quantum hoop conjecture: the de Broglie wavelength of a quantum system can not be arbitrarily small, it must be larger than the characterized Schwarzschild radius of the quantum system. This conjecture gives an upper bound for the wave number or the momentum of the quantum system. For application, we found a natural cutoff for the vacuum energy of a scalar field.

\begin{acknowledgments}
This study is supported in part by National Natural Science Foundation of China (Grant Nos. 11147028 and 11273010), Hebei Provincial Natural Science Foundation of China (Grant No. A2014201068), the Outstanding Youth Fund of Hebei University (No. 2012JQ02), the Open Project Program of State Key Laboratory of Theoretical Physics, Institute of Theoretical Physics, Chinese Academy of Sciences, China (No.Y4KF101CJ1), and the Midwest universities comprehensive strength promotion project.
\end{acknowledgments}

\bibliographystyle{elsarticle-num}
\bibliography{ref}

\end{document}